\documentclass[lettersize,journal]{IEEEtran}
\usepackage{amsmath,amsfonts}
\usepackage{algorithm}
\usepackage{array}
\usepackage[caption=false,font=normalsize,labelfont=sf,textfont=sf]{subfig}
\usepackage{textcomp}
\usepackage{stfloats}
\usepackage{url}
\usepackage{verbatim}
\usepackage{graphicx}
\usepackage{cite}
\usepackage{hyperref}
\usepackage{algpseudocode}
\usepackage{setspace}
\usepackage{booktabs}
\usepackage{ragged2e} 
\usepackage[normalem]{ulem}
\newcommand{\duline}[1]{\uuline{#1}}

\begin{document}

\title{ XVAE-WMT: Explainable Wavelet-Temporal Variational Autoencoder for Blind Source Separation of Heart and Lung Sounds }

\author{
Yasaman Torabi$^{1,\star}$, 
Shahram Shirani$^{1,2}$, 
James P.~Reilly$^{1}$
\\[0.5em]
\small
$^{1}$ Department of Electrical and Computer Engineering,
McMaster University, Hamilton, ON L8S 4K1, Canada.
\\
\small
$^{2}$ L.R. Wilson/Bell Canada Chair in Data Communications, Hamilton, ON L8S 4L7, Canada.
\\
\small
$\star$ Corresponding author: torabiy@mcmaster.ca
}

\maketitle

\begin{abstract}
The separation of cardiovascular sounds is a critical task in biomedical signal 
processing. In this paper, we introduce XVAE-WMT\footnote{ The python scripts are available on: \url{https://github.com/Torabiy/XVAE}}, an unsupervised explainable generative AI algorithm 
combining a variational autoencoder (VAE) with explainable AI (XAI), wavelet-based 
inputs, a post-hoc output mask, and temporal consistency (TC) loss. Unlike existing 
supervised and VAE-based methods that rely on Short-Time Fourier Transform (STFT) 
and ignore latent interpretability, XVAE-WMT requires no paired clean recordings 
and integrates a Continuous Wavelet Transform (CWT) front-end for superior 
time-frequency localization. We assessed the latent space 
interpretability via different metrics, with SHAP (SHapley Additive exPlanations) enabling dimensionality reduction to the top 75\% of latent features while preserving separation quality. Evaluated across two datasets using Signal-to-Distortion Ratio (SDR), Signal-to-Interference Ratio (SIR), and Signal-to-Artifacts Ratio (SAR), XVAE-WMT attains 26.8\,dB SDR, 32.8\,dB SIR, and 28.6\,dB SAR.
\end{abstract}

\section{Introduction}

\IEEEPARstart{T}{he} separation of heart and lung sounds from 
single-channel recordings is a fundamental challenge in 
biomedical acoustic signal processing, with direct 
implications for clinical diagnosis and patient monitoring 
\cite{ref1}, \cite{ref2}. In practice, cardiopulmonary 
sounds are inevitably mixed during auscultation, as the 
mechanical activity of the heart and the airflow dynamics 
of the lungs produce overlapping acoustic signatures that 
are difficult to disentangle \cite{ref3}. Accurate 
separation of these sources enables clinicians to assess 
cardiac and respiratory conditions independently \cite{gupta2020ae}. The challenge is further 
compounded by the non-stationary and cyclic nature of both 
signals, which limits the applicability of conventional 
signal processing approaches that assume stationarity or 
statistical independence among sources. Traditional blind source separation (BSS) methods such as 
Independent Component Analysis (ICA) and Non-Negative 
Matrix Factorization (NMF) have been widely applied to 
cardiopulmonary sound separation \cite{thesis}, \cite{ref4}. 
ICA assumes statistical independence among sources, a 
condition that is frequently violated in real-world 
recordings where heart and lung sounds exhibit overlapping 
frequency content and correlated temporal patterns 
\cite{hyvarinen2001ica}, \cite{comon1994ica}. NMF relies 
on non-negativity constraints and performs well on spectral 
data, but struggles to capture the temporal coherence 
inherent in cyclic physiological signals \cite{ref5}, 
\cite{lee2001algorithms}. Principal Component Analysis 
(PCA) seeks uncorrelated linear projections of the data 
but fails to achieve meaningful separation when source 
signals are non-orthogonal or exhibit nonlinear 
interactions \cite{sharma2016pca}. Empirical Mode 
Decomposition (EMD) decomposes signals into intrinsic mode 
functions but suffers from mode mixing and lacks robustness 
in noisy environments where heart and lung components 
overlap \cite{khodayari2013emd}. These fundamental 
limitations have motivated the development of more 
adaptive, data-driven approaches that do not rely on strict 
assumptions about source characteristics or require 
multiple microphones for spatial diversity.

Recent advances in deep learning and generative modeling 
have opened new avenues for single-channel source 
separation that overcome many of the limitations of 
classical methods \cite{mamba}. Autoencoders (AEs) have been widely 
adopted for biomedical signal denoising and compression, 
leveraging their ability to learn compact latent 
representations in an unsupervised manner \cite{vincent2008denoising}, 
\cite{gupta2020ae}, \cite{aaaaaLI2023110176}. The broader 
class of generative AI (GenAI) models, including 
Generative Adversarial Networks (GANs) and diffusion 
models, has demonstrated strong performance in audio 
synthesis and enhancement even under limited supervision 
\cite{yang2022ganvoice}. Among these, Variational 
Autoencoders (VAEs) are
well-suited for source separation tasks, as they learn 
structured probabilistic latent spaces that enable 
effective modeling of complex signal mixtures 
\cite{kingma2013auto}, \cite{Karamatli_2019}. Compared with a standard AE, the VAE learns a regularized probabilistic latent space rather than a deterministic representation, which facilitates smooth and structured latent representations for source-specific analysis. The stochastic sampling also enables the model to capture uncertainty in the latent representation while maintaining differentiable training through the reparameterization trick. VAE-based 
models have been successfully applied to blind source 
separation in speech, music, and biomedical signals, 
demonstrating improvements over classical 
baselines \cite{pandey2018vae}, \cite{leglaive2020conditional}, 
\cite{e24010055}. Despite these advances, existing VAE-based separation 
models do not fully exploit the temporal structure of 
cardiopulmonary signals, nor do they incorporate 
time-frequency representations suited to non-stationary 
biomedical data. 

Temporal consistency (TC) loss, which 
penalizes abrupt transitions between consecutive time 
steps, has been shown to improve sequence coherence in 
models applied to cyclic physiological data \cite{TC}. Wavelet transforms provide a compact and multi-resolution time-frequency 
representation that outperforms the Short-Time Fourier 
Transform (STFT) for non-stationary signals such as 
heartbeats and breath sounds, offering superior 
localization in both time and frequency domains 
\cite{mallat2008wavelet}, \cite{daubechies1992ten}, 
\cite{kazemi2017wavelet}. Furthermore, mask-based 
post-processing has been introduced in generative 
separation models to improve output fidelity by applying 
an element-wise ratio between the original mixture and the 
reconstructed signal, enforcing consistency while 
preserving fine spectral detail \cite{kong2024audioMAE}, 
\cite{zhao2024vaeJCR}. 

As deep learning models grow in complexity, 
interpretability has become an equally important concern, 
particularly in biomedical applications where transparency 
is essential for clinical trust and regulatory compliance 
\cite{holzinger2023fusion}. Explainable AI (XAI) methods 
have been applied to analyze latent space structure in 
generative models \cite{molnar2021book}. SHAP 
(SHapley Additive exPlanations) has emerged as a 
particularly reliable approach for quantifying the 
contribution of individual features to model outputs \cite{lundberg2017shap}, \cite{zhang2022xdeep}.

In this paper, we propose XVAE-WMT, a novel explainable 
generative framework for unsupervised blind source 
separation of heart and lung sounds. The model operates in 
a single-microphone setting and integrates three key 
components into the VAE framework: wavelet-based 
time-frequency inputs (W), a post-hoc output mask (M), 
and temporal consistency loss (T). The main 
contributions of this work are as follows:

\begin{itemize}
    \item We propose XVAE-WMT, a VAE-based model 
    integrating XAI designed for blind source separation of cardiopulmonary sounds.
    \item We incorporate wavelet-based inputs into the VAE 
    framework, demonstrating superior time-frequency 
    efficiency over STFT representations for non-stationary signals.
    \item We introduce temporal consistency loss into the 
    VAE objective to enforce smooth transitions and preserve 
    the cyclic structure of heart and lung sounds.
    \item We apply a post-hoc output mask to normalize 
    estimated sources relative to the original mixture.
    \item We provide a latent space 
    interpretability analysis using clustering metrics and 
    SHAP, enabling dimensionality reduction 
    without sacrificing separation performance.
\end{itemize}

\begin{figure*}
    \centering
    \includegraphics[width=0.9\linewidth, trim={0 4cm 0 3cm}, clip]{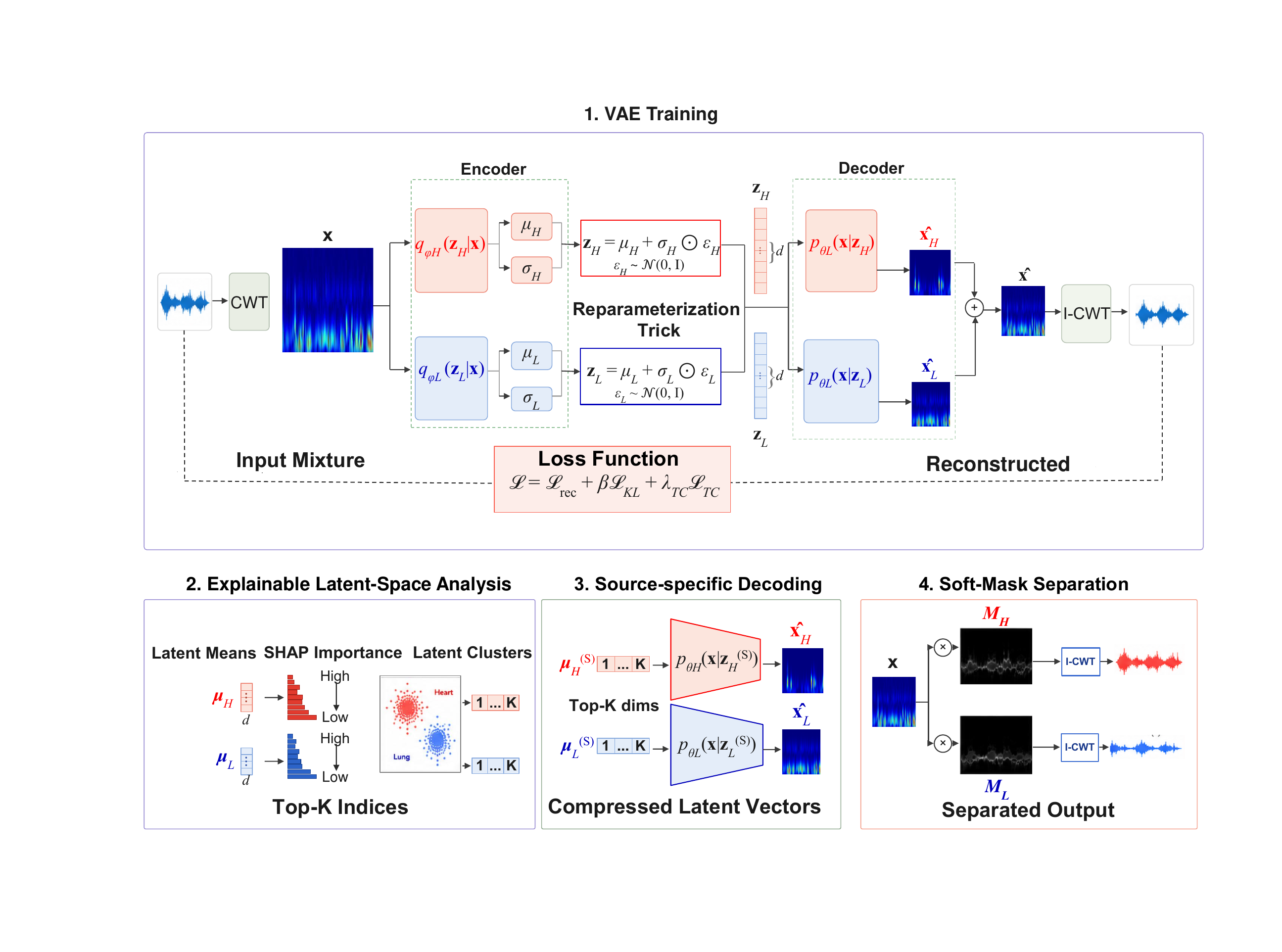}
    \captionsetup{justification=justified,font=footnotesize}
    \caption{XVAE-WMT framework: (1) VAE Training, where the mixture is transformed into a wavelet scalogram and encoded into latent representations; (2) Explainable Latent-Space Analysis, where SHAP identifies the most informative latent dimensions and clustering reveals source-related structure; (3) Source-specific Decoding, where the selected latent dimensions are decoded into heart and lung representations; and (4) Soft-Mask Separation, where source-specific masks are applied to the mixture scalogram and inverse wavelet transformation produces the separated heart and lung signals.}
    \label{fig:flow}
\end{figure*}

\section{Related Work}
\label{sec:related}

Early approaches to cardiopulmonary sound separation 
relied on ICA and NMF 
\cite{chem}. While ICA exploits statistical 
independence between sources, this assumption is frequently 
violated in clinical recordings where heart and lung sounds 
share overlapping spectral content \cite{hyvarinen2001ica}, 
\cite{comon1994ica}. NMF offers better spectral 
decomposition without strict independence requirements, 
though it struggles with temporal dynamics \cite{ref4}. 
Periodicity-constrained NMF extensions have been proposed 
specifically for cardiopulmonary signals to exploit the 
rhythmic structure of heartbeats and breathing cycles 
\cite{ref6}, \cite{lee2001algorithms}. Despite these 
refinements, classical methods remain limited by their 
inability to capture the nonlinear and non-stationary 
characteristics of real cardiopulmonary recordings, 
motivating the shift toward data-driven approaches. Deep learning has significantly advanced biomedical source 
separation over the past decade. Autoencoder-based models 
improved unsupervised representation learning for 
source-specific feature extraction \cite{vincent2008denoising}, 
\cite{aaaaaLI2023110176}, with Tsai et al. \cite{gupta2020ae} 
proposing a periodicity-coded deep autoencoder (PC-DAE) 
that incorporates structural priors about cardiopulmonary 
periodicity. Supervised architectures such as LSTM networks 
and U-Net variants achieve strong performance but require 
large labeled datasets of clean recordings that are 
difficult to obtain clinically \cite{bai2018empirical}. 
GAN-based generative models have demonstrated strong audio 
enhancement performance even under weak supervision 
\cite{yang2022ganvoice}, motivating the exploration of 
fully unsupervised generative frameworks that do not depend 
on paired clean recordings.

VAEs are considered a principled framework for unsupervised 
source separation, regularizing the latent space to enable both reconstruction and generative 
sampling \cite{kingma2013auto}, \cite{Karamatli_2019}. 
Monaural VAEs have demonstrated effective separation of 
overlapping speech signals \cite{pandey2018vae}, while 
conditional extensions incorporate class-level priors for 
source-specific generation \cite{leglaive2020conditional}. 
Architectural variants including VQ-VAE \cite{oord2017vqvae} 
and RQ-VAE \cite{berti2024rqvae} further extend generative 
capacity through latent space discretization and 
hierarchical quantization \cite{e24010055}. However, 
existing VAE-based models have not fully exploited the 
temporal structure and non-stationary time-frequency 
characteristics specific to biomedical audio. Temporal consistency constraints have been applied in 
sequence modeling to enforce smooth transitions between 
consecutive predictions \cite{TC}. Wavelet-based representations provide multi-resolution 
time-frequency decomposition better suited to non-stationary 
signals than STFT \cite{mallat2008wavelet}, 
\cite{daubechies1992ten}, and have been shown to improve 
VAE separation performance for transient biomedical signals 
\cite{kazemi2017wavelet}. Mask-based post-processing has 
further been proposed to refine generative model outputs 
by enforcing consistency with the original mixture 
\cite{kong2024audioMAE}, \cite{zhao2024vaeJCR}. On the 
interpretability front, latent space clustering metrics 
and SHAP-based feature attribution have been applied to 
generative models to quantify source separability and 
guide dimensionality reduction \cite{molnar2021book}, 
\cite{lundberg2017shap}, \cite{zhang2022xdeep}. Despite these advances, XAI has not been integrated into VAE-based blind source separation for biomedical audio, a gap that the proposed XVAE-WMT framework directly addresses.

\section{Theoretical Background}
\begin{figure}
    \centering
    \includegraphics[width=\linewidth, trim={0 4cm 0 6cm}, clip]{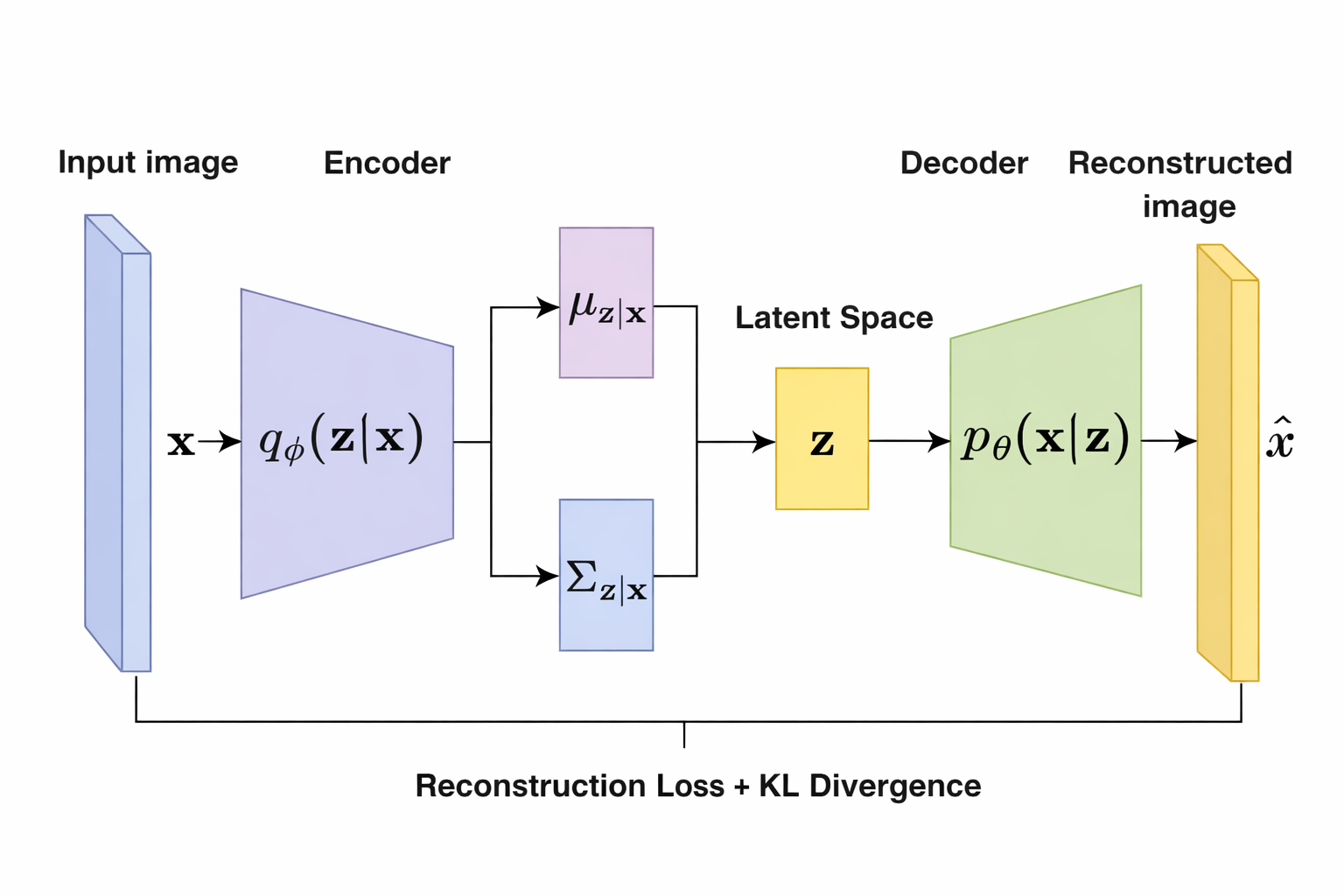}
    \captionsetup{justification=justified,font=footnotesize}
    \caption{Schematic overview of the VAE architecture. The encoder $q_\phi(z \mid x)$ maps the input $x$ into a probabilistic latent representation parameterized by the mean $\mu_{z\mid x}$ and covariance $\Sigma_{z\mid x}$. 
    A latent variable $z$ is sampled from this distribution and passed to the decoder $p_\theta(x \mid z)$ to reconstruct the input signal $\hat{x}$. 
    Training is performed by minimizing the reconstruction loss together with the Kullback--Leibler divergence(KL). The distributions $q_\phi(\cdot\mid\mathbf{x})$ and $p_\theta(\cdot\mid\mathbf{z})$ are parameterized by neural networks trained jointly via variational inference.
}
    \label{fig:vae_arch}
\end{figure}

Autoencoders are unsupervised neural models that learn to reconstruct their input through a latent bottleneck representation, and consist of an encoder and a decoder. Variational Autoencoders (VAEs) are a probabilistic extension of autoencoders. While standard autoencoders learn a deterministic mapping to a latent bottleneck, VAEs learn a distribution over the latent space. Specifically, the encoder maps an input to the parameters of a probability distribution, and the decoder reconstructs the input by sampling from this distribution. This enables VAEs not only to reconstruct inputs but also to generate new data by sampling from the learned latent space. Figure~\ref{fig:vae_arch} illustrates the overall architecture of a variational autoencoder (VAE). 

In a standard autoencoder, the encoder maps an input vector \( \mathbf{x} \in \mathbb{R}^n \) to a latent vector \( \mathbf{h} \in \mathbb{R}^k \) (\ref{eq:encoder}), and the decoder reconstructs the input from this latent representation (\ref{eq:decoder}). The goal of training is to minimize the reconstruction loss over the data distribution of $p_{data}$, as shown in (\ref{eq:ae_loss}). Here, $\Delta$ denotes a reconstruction loss function measuring the discrepancy between the input and its reconstruction. $\Delta$ is typically chosen as the mean squared error (MSE).
\begin{equation}
\mathbf{h} = g(\mathbf{x}).
\label{eq:encoder}
\end{equation}
\begin{equation}
\tilde{\mathbf{x}} = f(\mathbf{h}) = f(g(\mathbf{x})).
\label{eq:decoder}
\end{equation}
\begin{equation}
\arg\min_{f,g}\;
\mathbb{E}_{\mathbf{x}\sim p_{\mathrm{data}}}
\left[\Delta\!\left(\mathbf{x},f(g(\mathbf{x}))\right)\right].
\label{eq:ae_loss}
\end{equation}

Here, the latent dimension satisfies $k \leq n$, enforcing a compressed representation of the input. The encoder $g$ and decoder $f$ are implemented as parameterized neural networks, where the encoder maps the input signal to a low-dimensional latent space and the decoder reconstructs the input from this latent representation. Although the autoencoder learns a compact latent representation, it cannot generate new data because it learns a deterministic mapping. Variational autoencoders (VAEs) address this limitation through a probabilistic framework that enables both reconstruction and generation. The encoder maps the input \( \mathbf{x} \) not to a fixed point but to a distribution in the latent space. The decoder reconstructs data by sampling from this distribution. We formulate the optimization problem as a maximum likelihood estimation (MLE) task, where the goal is to maximize the log-likelihood of the observed data under the model parameters:
\begin{equation}
\max_{\theta}\,\log p_{\theta}(\mathbf{x})
\label{eq:vae_mle_objective}
\end{equation}
We start from the marginal data likelihood:
\begin{equation}
\log p_{\theta}(\mathbf{x})=\log \int p_{\theta}(\mathbf{x},\mathbf{z})\,d\mathbf{z}.
\label{eq:vae_loglike}
\end{equation}
Here, $z \in \mathbb{R}^k$ denotes the latent variable associated with the bottleneck of the autoencoder, representing the low-dimensional latent representation of the input signal. Since the integral in (\ref{eq:vae_loglike}) is generally intractable, we introduce an approximate posterior \(q_{\phi}(\mathbf{z}\,|\,\mathbf{x})\) and rewrite the expression by multiplying and dividing inside the integral:
\begin{align}
\log p_{\theta}(\mathbf{x})
&=\log \int \frac{p_{\theta}(\mathbf{x},\mathbf{z})}{q_{\phi}(\mathbf{z}\,|\,\mathbf{x})}\,q_{\phi}(\mathbf{z}\,|\,\mathbf{x})\,d\mathbf{z} \notag\\
&=\log \mathbb{E}_{\mathbf{z}\sim q_{\phi}}\!\left[\frac{p_{\theta}(\mathbf{x},\mathbf{z})}{q_{\phi}(\mathbf{z}\,|\,\mathbf{x})}\right].
\label{eq:vae_expectation}
\end{align}
Applying Jensen’s inequality yields a lower bound on the log-likelihood:
\begin{equation}
\log p_{\theta}(\mathbf{x})
\geq \mathbb{E}_{\mathbf{z}\sim q_{\phi}}\!\left[\log \frac{p_{\theta}(\mathbf{x},\mathbf{z})}{q_{\phi}(\mathbf{z}\,|\,\mathbf{x})}\right].
\label{eq:elbo_inequality}
\end{equation}
Expanding the joint distribution in (\ref{eq:elbo_inequality}) and separating the terms inside the logarithm, we obtain:
\begin{align}
\log \frac{p_{\theta}(\mathbf{x},\mathbf{z})}{q_{\phi}(\mathbf{z}\,|\,\mathbf{x})}=\log p_{\theta}(\mathbf{x}\,|\,\mathbf{z}) +\log\!\frac{p_{\theta}(\mathbf{z})}{q_{\phi}(\mathbf{z}\,|\,\mathbf{x})}.
\label{eq:elbo_expand_raw}
\end{align}
For two continuous probability distributions \(a(\mathbf{x})\) and \(b(\mathbf{x})\), the Kullback–Leibler (KL) divergence is defined as
\begin{equation}
\mathcal{D}_{\mathrm{KL}}(a\parallel b)
=\mathbb{E}_{\mathbf{x}\sim a}\!\left[\log\!\frac{a(\mathbf{x})}{b(\mathbf{x})}\right]
=\int a(\mathbf{x})\log\!\frac{a(\mathbf{x})}{b(\mathbf{x})}\,d\mathbf{x}.
\label{eq:kl_def}
\end{equation}
The second expectation term in (\ref{eq:elbo_expand_raw}) corresponds to the negative KL divergence between the approximate posterior \(q_{\phi}(\mathbf{z}\,|\,\mathbf{x})\) and the prior \(p_{\theta}(\mathbf{z})\):
\begin{equation}
\mathbb{E}_{\mathbf{z}\sim q_{\phi}}\!\left[\log\!\frac{p_{\theta}(\mathbf{z})}{q_{\phi}(\mathbf{z}\,|\,\mathbf{x})}\right]
=-\,\mathcal{D}_{\mathrm{KL}}\!\big(q_{\phi}(\mathbf{z}\,|\,\mathbf{x})\,\|\,p_{\theta}(\mathbf{z})\big).
\label{eq:elbo_klterm}
\end{equation}
Substituting (\ref{eq:elbo_klterm}) into (\ref{eq:elbo_expand_raw}) yields the \emph{Evidence Lower Bound} (ELBO), which serves as the objective function optimized during VAE training:
\begin{equation}
\mathrm{ELBO}
=
-\mathcal{D}_{\mathrm{KL}}\!\big(q_{\phi}(\mathbf{z}\,|\,\mathbf{x})\,\|\,p_{\theta}(\mathbf{z})\big)
+\mathbb{E}_{\mathbf{z}\sim q_{\phi}}[\log p_{\theta}(\mathbf{x}\,|\,\mathbf{z})].
\label{eq:elbo}
\end{equation}
The ELBO is maximized with respect to the model parameters $\theta$ and $\phi$, where $\theta$ denotes the parameters of the decoder network $p_\theta(x|z)$ and $\phi$ denotes the parameters of the encoder network $q_\phi(z|x)$. Maximizing this bound regularizes the latent space to match the prior in the encoder through the first KL regularization term, and it encourages accurate reconstruction of the input in the decoder through the second likelihood term. Given the assumption that the prior distribution follows a standard multivariate 
normal $p(\mathbf{z}) \equiv \mathcal{N}(\mathbf{0}, \mathbf{I})$, and that the 
approximate posterior 
$q_{\phi}(\mathbf{z}\,|\,\mathbf{x}) = \mathcal{N}(\boldsymbol{\mu}, 
\mathrm{diag}(\boldsymbol{\sigma}^2))$ 
is Gaussian with mean vector $\boldsymbol{\mu} \in \mathbb{R}^k$ and standard 
deviation vector $\boldsymbol{\sigma} \in \mathbb{R}^k$, 
the ELBO in (\ref{eq:elbo}) simplifies to a closed-form expression. 
The KL divergence term becomes
\begin{equation}
\mathcal{D}_{\mathrm{KL}}\!\big(q_{\phi}(\mathbf{z}\,|\,\mathbf{x})\,\|\,p_{\theta}(\mathbf{z})\big)
=\tfrac{1}{2}\sum_{i=1}^{k}\!\left(\sigma_{i}^{2}+\mu_{i}^{2}-1-\log\sigma_{i}^{2}\right),
\label{eq:kl_gauss_simplified}
\end{equation}
which regularizes the latent representation to remain close to the unit Gaussian prior. The closed-form KL expression does not replace the encoder. It uses the $\boldsymbol{\mu}$ and $\boldsymbol{\sigma}$ produced by the encoder for each input to evaluate the divergence analytically. In addition, the reconstruction term $\mathbb{E}_{\mathbf{z}\sim q_{\phi}}[\log p_{\theta}(\mathbf{x}\,|\,\mathbf{z})]$ can be 
simplified under the same Gaussian assumption and  implemented as the mean squared error 
(MSE) loss between the input mixture and its reconstruction.

\section{Methodology}

The proposed XVAE-WMT method integrates a wavelet-based time--frequency front-end, a variational autoencoder (VAE) with a temporal consistency (TC) regularizer, and a post-hoc SHAP-based latent analysis. An overview of the full procedure is given in Algorithm~\ref{alg:xvae_wmt}. During training, each mixture is converted to a wavelet scalogram and passed through the encoder to obtain $\boldsymbol{\mu}$ and $\boldsymbol{\sigma}$, from which a latent sample is generated and decoded. The reconstruction, KL, and temporal-consistency losses are then jointly optimized. After training, a new mixture is transformed into a scalogram, encoded into the latent space, and separated by decoding the source-specific latent representations, constructing soft masks, and applying the masks followed by inverse CWT to obtain the heart and lung signals.

\subsection{Wavelet Front-End} 
Each input mixture $m \in \mathbb{R}^{T}$ is transformed via the Continuous Wavelet Transform (CWT) using a mother wavelet $\psi$ and a scale set $\mathcal{S}$ of size $F$, producing a non-negative scalogram $x = |\mathrm{CWT}(m;\psi,\mathcal{S})| \in \mathbb{R}^{F \times T}$. Unlike the Short-Time Fourier Transform, the CWT provides multi-resolution time--frequency localization, which is well-suited to the non-stationary and quasi-periodic nature of heart and lung sounds. CWT is also purely real. Recent studies have also demonstrated that physiological responses to external stimulation can exhibit frequency-dependent dynamics, further highlighting the importance of frequency-aware analysis in biomedical signals \cite{s1,s2}. It is worth mentioning that the term "Continuous" refers to how the math smoothly scans a signal. Even with digital data, CWT creates a highly detailed 2D frequency map by shifting the wavelet sample-by-sample along a grid. In contrast, the Discrete Wavelet Transform (DWT) skips most frequencies by doubling its steps, making DWT best for data compression and CWT the choice for signal processing.
\subsection{Encoder-Decoder Structure} 
The VAE consists of an encoder network parameterizing the posterior distribution $q_\phi(\mathbf{z}\mid\mathbf{x})$ and a decoder network parameterizing the likelihood distribution $p_\theta(\mathbf{x}\mid\mathbf{z})$. The encoder maps the scalogram $x$ to the parameters of a Gaussian posterior over the latent variable $z \in \mathbb{R}^{d}$, i.e., the mean vector $\boldsymbol{\mu} \in \mathbb{R}^{d}$ and log-variance $\log\boldsymbol{\sigma}^{2} \in \mathbb{R}^{d}$. In our implementation, $d = 128$. The encoder comprises three convolutional blocks: (i) 128 filters with a $1 \times F$ kernel to capture spectral patterns across scales, (ii) 128 filters with a $4 \times 1$ kernel to model short-range temporal dependencies, and (iii) 256 filters with a $4 \times 1$ kernel to extract higher-level spatio-temporal correlations. The resulting feature map is flattened and passed through two fully connected layers that output $\boldsymbol{\mu}$ and $\log\boldsymbol{\sigma}^{2}$. The decoder mirrors this structure with transposed convolutions and reconstructs a scalogram $\hat{x} \in \mathbb{R}^{F \times T}$. Latent samples are drawn using the reparameterization trick \cite{kingma2013auto}, which expresses the stochastic sampling as a differentiable function of deterministic network outputs, enabling end-to-end gradient-based training. 
\begin{equation}
z = \boldsymbol{\mu} + \boldsymbol{\sigma} \odot \boldsymbol{\epsilon}, \qquad \boldsymbol{\epsilon} \sim \mathcal{N}(\mathbf{0}, \mathbf{I}_d),
\end{equation}
Here, $\mu$ and $\sigma$ are obtained as the outputs of the encoder network $q_\phi(z|x)$ for each input $x$. The reparameterization trick separates the random sampling from the network parameters, allowing gradients to propagate through the stochastic latent representation during training.

\subsection{Training Objective} 
Training maximizes the Evidence Lower Bound (ELBO) of the VAE with respect to $\phi$ and $ \theta$, augmented by a temporal consistency term. Under a standard Gaussian prior $p(z) = \mathcal{N}(\mathbf{0}, \mathbf{I}_d)$ and a Gaussian likelihood, the ELBO reduces to a reconstruction (MSE) term and a closed-form KL term:
\begin{align}
\mathcal{L}_{\text{rec}} &= \| x - \hat{x} \|_2^{2}, \\
\mathcal{L}_{\text{KL}}  &= \tfrac{1}{2}\sum_{k=1}^{d}\!\big(\sigma_k^{2} + \mu_k^{2} - 1 - \log\sigma_k^{2}\big).
\end{align}

To suppress spurious frame-to-frame fluctuations in the reconstructed scalogram, we further introduce a temporal consistency regularizer
\begin{equation}
\mathcal{L}_{\text{TC}} = \sum_{t} \| \hat{x}_{t+1} - \hat{x}_{t} \|_2^{2},
\end{equation}
which penalizes abrupt variations between adjacent time frames and encourages physiologically plausible, smoothly evolving reconstructions. The total training loss is
\begin{equation}
\mathcal{L} = \mathcal{L}_{\text{rec}} + \beta\,\mathcal{L}_{\text{KL}} + \lambda_{\text{TC}}\,\mathcal{L}_{\text{TC}},
\end{equation}
where $\beta$ controls the strength of latent regularization and $\lambda_{\text{TC}}$ balances temporal smoothness against reconstruction fidelity. Parameters $\phi$ and $\theta$ are updated jointly via stochastic gradient descent.
\subsection{Explainable Latent-Space Analysis} 
After training, we interpret the learned latent space to identify which dimensions of $\boldsymbol{\mu}$ are most informative for each source. We first visualize the latent representations using t-SNE, then perform a SHAP analysis to assign an importance score $|\varphi_k|$ to every latent dimension $k \in \{1,\dots,d\}$. The top-$K$ dimensions most strongly associated with each source based on the post-hoc SHAP analysis are retained to form the source-specific latent vectors $\boldsymbol{\mu}_{\text{H}}, \boldsymbol{\mu}_{\text{L}} \in \mathbb{R}^{d}$, corresponding to heart, lung sources, respectively. This procedure both explains the model's internal representation and acts as a principled, data-driven dimensionality reduction. Importantly, no heart or lung labels are used during VAE training. The source labels are used only post hoc to interpret the learned latent clusters and associate the two latent groups with heart and lung sources.

\subsection{Post-hoc Soft-Mask Separation} 
Unlike a conventional classifier that assigns each input to a single class, our method performs source separation by estimating a continuous contribution of each source at every time--frequency bin. The soft mask therefore preserves overlapping heart and lung components rather than making a hard class decision. Separation is performed at inference time by decoding each source-specific latent vector and constructing a soft time--frequency mask. For each source $c \in \{\text{H}, \text{L}\}$,
\begin{equation}
\hat{x}_c = p_\theta(\boldsymbol{\mu}_c), \qquad
M_c(t,f) = \frac{\hat{x}_c(t,f)}{\hat{x}_{\text{H}}(t,f) + \hat{x}_{\text{L}}(t,f)},
\end{equation}
so that the masks partition the mixture energy across sources at every time--frequency bin and satisfy $M_{\text{H}} + M_{\text{L}} = 1$. Each mask is applied element-wise to the mixture scalogram, and the corresponding time-domain source estimate is obtained via the inverse CWT:
\begin{equation}
\hat{s}_c = \mathrm{iCWT}\!\left(M_c \odot x;\, \psi, \mathcal{S}\right) \in \mathbb{R}^{T}.
\end{equation}

\begin{algorithm}[H]
\footnotesize
\caption{XVAE-WMT}\label{alg:xvae_wmt}
\setstretch{1.2}
\begin{algorithmic}[1]
\Statex \textbf{Input:} mixture $m \in \mathbb{R}^{T}$; mother wavelet $\psi$; scale set $\mathcal{S} \in \mathbb{R}^{F}$; weights $\lambda_{\text{TC}}, \beta \in \mathbb{R}$; top-$K \in \mathbb{N}$
\Statex \textbf{Output:} separated sources $\hat{s}_{\text{H}}, \hat{s}_{\text{L}} \in \mathbb{R}^{T}$
\Function{$q_\phi$}{$x \in \mathbb{R}^{F \times T}$}
    \State Pass $x$ through the encoder NN with weights $\phi$
    \State \Return $(\boldsymbol{\mu},\, \log\boldsymbol{\sigma}^{2}) \in \mathbb{R}^{d} \times \mathbb{R}^{d}$
\EndFunction
\Function{$p_\theta$}{$z \in \mathbb{R}^{d}$}
    \State Pass $z$ through the decoder NN with weights $\theta$
    \State \Return $\hat{x} \in \mathbb{R}^{F \times T}$
\EndFunction
\State Initialize parameters $\phi, \theta$
\State $x \gets |\mathrm{CWT}(m;\, \psi, \mathcal{S})| \in \mathbb{R}^{F \times T}$
\For{each mini-batch}
    \State $(\boldsymbol{\mu},\, \log\boldsymbol{\sigma}^{2}) \gets \Call{$q_\phi$}{x}$
    \State Sample $\boldsymbol{\epsilon} \sim \mathcal{N}(\mathbf{0}, \mathbf{I}_d)$
    \State $z \gets \boldsymbol{\mu} + \boldsymbol{\sigma} \odot \boldsymbol{\epsilon}$
    \State $\hat{x} \gets \Call{$p_\theta$}{z}$
    \State $\mathcal{L}_{\text{rec}} = \| x - \hat{x} \|_2^2$
    \State $\mathcal{L}_{\text{KL}} = \tfrac{1}{2}\sum_{k=1}^{d}\!\big(\sigma_k^{2} + \mu_k^{2} - 1 - \log\sigma_k^{2}\big)$
    \State $\mathcal{L}_{\text{TC}} = \sum_{t} \| \hat{x}_{t+1} - \hat{x}_{t} \|_2^2$
    \State $\mathcal{L} = \mathcal{L}_{\text{rec}} + \beta\,\mathcal{L}_{\text{KL}} + \lambda_{\text{TC}}\,\mathcal{L}_{\text{TC}}$
    \State Update $\theta, \phi$ via $\nabla \mathcal{L}$
\EndFor
\State Rank latent dimensions of $\boldsymbol{\mu}$ by SHAP, and select the top-$K$ to obtain $\boldsymbol{\mu}_{\text{H}}, \boldsymbol{\mu}_{\text{L}} \in \mathbb{R}^{d}$
\For{$c \in \{\text{H}, \text{L}\}$}
    \State $\hat{x}_c \gets \Call{$p_\theta$}{\boldsymbol{\mu}_c}$
    \State $M_c \gets \hat{x}_c / (\hat{x}_{\text{H}} + \hat{x}_{\text{L}})$
    \State $\hat{s}_c \gets \mathrm{iCWT}\!\left(M_c \odot x;\, \psi, \mathcal{S}\right)$
\EndFor
\end{algorithmic}
\end{algorithm}
\footnotesize\textbf{\textit{Notes:}} NN: Neural Network,   iCWT: inverse Continuous Wavelet Transform, SHAP: SHapley Additive exPlanations for feature attribution.

\normalsize
\section{Experimental Procedure}

\subsection{Dataset}
We trained the model on two datasets. We created the first dataset by mixing the Kaggle Respiratory Sound Database, and the CirCor DigiScope Heart Sound Database. All recordings were resampled to 16 kHz, normalized, and segmented into 1-second frames, producing 24,383 heart and 18,144 lung segments across training iterations. Random pairs were combined to generate the mixtures. We collected the second dataset (HLS-CMDS) from a clinical manikin\cite{Torabi2024}. After resampling, normalization, and segmentation, we randomly paired heart and lung segments, forming 25,000 mixtures. Each sound frame was transformed into a wavelet spectrogram using the Mexican Hat mother wavelet.

\subsection{Network Architecture}
The model follows a convolutional encoder--decoder structure operating on time--frequency representations of size $T \times F$, where $T$ and $F$ denote the time and frequency axes, respectively. Two front-ends are considered: a Short-Time Fourier Transform (STFT) with $F=256$ frequency bins, and a Continuous Wavelet Transform (CWT) with $F=200$ scales. Both produce magnitude spectrograms that serve as inputs to the network. Each convolutional layer is followed by batch normalization and a ReLU activation, except for the final layer. The encoder outputs the posterior mean $\boldsymbol{\mu}$ and log-variance $\log\boldsymbol{\sigma}^{2}$ of the latent variable $z \in \mathbb{R}^{128}$. The decoder terminates with a sigmoid activation to produce a soft time--frequency mask. The encoder architecture is summarized in Table~\ref{tab:hidden_units}. The decoder mirrors the encoder using transposed convolutions. 

\begin{table}[h]
\centering
\scriptsize
\captionsetup{font=footnotesize}
\caption{Encoder network architecture.}
\vspace{-0.25 cm}
\label{tab:hidden_units}
\begin{tabular}{lccc}
\toprule
\textbf{Layer} & \textbf{Filters} & \textbf{Kernel} & \textbf{Stride} \\
\midrule
Conv 1   & 128 & $1\times F$ & $1\times1$ \\
Conv 2   & 128 & $4\times1$  & $2\times1$ \\
Conv 3   & 256 & $4\times1$  & $2\times1$ \\
FC       & 512 & --          & --         \\
Output   & 128 & --          & --         \\
\bottomrule
\end{tabular}
\end{table}

\vspace{-0.65 cm}
\subsection{Evaluation Metrics}
We evaluate the proposed method using different criteria:

\subsubsection{Separation Quality}
For $N$ source signals $s_i(t)$ producing a mixture, the goal is to recover an estimate $\hat{s}_i(t)$ of each $s_i(t)$. Each estimate can be decomposed as
\begin{equation}
\hat{s}(t) = s_{\text{target}}(t) + e_{\text{interf}}(t) + e_{\text{noise}}(t) + e_{\text{artif}}(t),
\end{equation}
where $s_{\text{target}}$ is the target contribution, $e_{\text{interf}}$ is interference from unwanted sources, $e_{\text{noise}}$ is the noise contribution, and $e_{\text{artif}}$ accounts for algorithmic artifacts. The standard BSS metrics are:
\begin{align}
\mathrm{SDR} &= 10\log_{10}\!\left(\frac{\|s_{\text{target}}\|^{2}}{\|e_{\text{interf}}+e_{\text{noise}}+e_{\text{artif}}\|^{2}}\right),\\[6pt]
\mathrm{SIR} &= 10\log_{10}\!\left(\frac{\|s_{\text{target}}\|^{2}}{\|e_{\text{interf}}\|^{2}}\right),\\[6pt]
\mathrm{SAR} &= 10\log_{10}\!\left(\frac{\|s_{\text{target}}+e_{\text{interf}}+e_{\text{noise}}\|^{2}}{\|e_{\text{artif}}\|^{2}}\right).
\end{align}
To summarize separation quality in a single number, we define the Mean Separation Quality:
\begin{equation}
\mathrm{MSQ} = \tfrac{1}{3}(\mathrm{SDR}+\mathrm{SIR}+\mathrm{SAR}).
\end{equation}

Two efficiency measures are defined to account for computation and representation cost jointly:
\begin{equation}
\begin{aligned}
\mathrm{TEM} &= \frac{\mathrm{MSQ}}{\text{average runtime (s)}},
\\[6pt]
\mathrm{CEM} &= \frac{\mathrm{MSQ}}{\text{spectral resolution units}},
\end{aligned}
\end{equation}
where TEM is the Time-Efficiency Metric and CEM is the Compression-Efficiency Metric. Spectral resolution is 256 for STFT and 200 for the wavelet front-end.

\subsubsection{Latent-Space Clustering Metrics}
We assess the quality of the learned latent space using four unsupervised clustering metrics applied to the posterior means $\boldsymbol{\mu}$:
\textit{Silhouette score} (Silh., $\uparrow$), \textit{Davies--Bouldin index} (DB, $\downarrow$), \textit{Calinski--Harabasz index} (CH, $\uparrow$), and the variance (Var, $\downarrow$).

\subsubsection{Interpretability Metrics}
We quantify the interpretability of the latent representation through four complementary measures. Let $\mathcal{A}$, $\mathcal{S}$, $\mathcal{P}$, $\mathcal{D}$, and $\mathcal{C}$ denote accuracy, stability, purity, diversity, and the composite interpretability score, respectively. Latent vectors are clustered using KMeans ($k=2$) and compared to ground-truth labels. Accuracy is the number of correct assignments per total samples, while stability measures the consistency of SHAP-based importance scores across repeated runs and input perturbations.
\begin{equation}
\mathcal{S} = 1 - \sigma(\text{scores}),
\end{equation}
where $\sigma$ denotes the standard deviation of explanation scores. Since $\sigma(\text{scores})\in(0,1)$, $|1-\sigma(\text{scores})|\leq 1$. Purity evaluates whether each latent cluster corresponds to a single source,
\begin{equation}
\mathcal{P} = \frac{1}{N} \sum_{k} \max_{j} |C_{k} \cap L_{j}|,
\end{equation}
where $C_{k}$ is the $k$-th cluster, $L_{j}$ is the ground-truth label set for class $j$, and $|C_{k}\cap L_{j}|$ counts samples in cluster $k$ belonging to class $j$. Diversity measures how broadly latent dimensions capture distinct patterns and is computed as the entropy of normalized SHAP importances,
\begin{equation}
\mathcal{D} = -\sum_{i} p_{i}\log p_{i},
\end{equation}
where $p_{i}$ is the normalized importance of latent dimension $i$. Finally, the composite interpretability score aggregates all measures as
\begin{equation}
\mathcal{C} = \tfrac{1}{4}(\mathcal{A} + \mathcal{S} + \mathcal{P} + \mathcal{D}).
\end{equation}

\section{Results}
\begin{figure}[H]
    \centering
    \includegraphics[width=\linewidth]{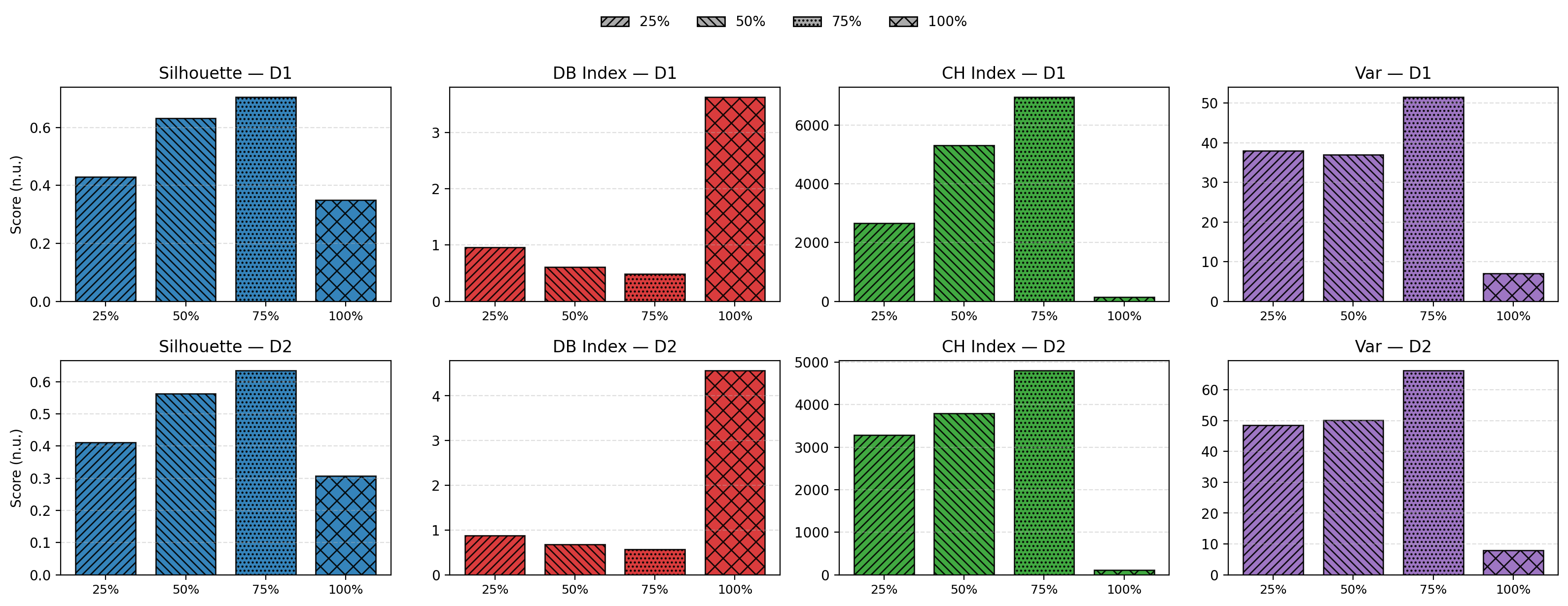}
    \captionsetup{justification=justified,font=footnotesize}
\caption{Clustering performance across different top-$k$\% subsets of the latent dimensions. Each bar represents the mean score for Silhouette, Davies–Bouldin (DB), Calinski–Harabasz (CH), and variance metrics 
on datasets D1 and D2. Higher Silhouette and CH, and lower DB and variance values indicate better clustering quality.}

    \label{fig:latent}
\end{figure}

We compare against two groups of baselines. The first group evaluates the contribution of each proposed component. Wavelet front-end (W), output masking (M), and temporal-consistency regularization (T) contribute a measurable improvement.  The second group benchmarks the proposed method against non-VAE separation approaches. Table~\ref{tab:results_all} reports evaluation metrics across all XVAE variants on both datasets. Table ~\ref{tab:latent_metrics} compares the latent-space clustering performance across XVAE variants. The results show some variation across metrics and datasets, with XVAE-WMT showing competitive clustering performance, particularly on Dataset One. We applied a SHAP-based explainability analysis on the latent embeddings to estimate feature importance and rank dimensions according to their contribution to clustering separability. We evaluated clustering performance for different top-k\% of the latent dimension (Fig.~\ref{fig:latent}). We retained the top 75\% of latent dimensions for final visualization and quantitative analysis, which yielded the most interpretable structure without degrading performance. 

\begin{table}[H]
\centering
\scriptsize
\setlength{\tabcolsep}{3pt}
\renewcommand{\arraystretch}{1.1}
\captionsetup{font=footnotesize}
\caption{Comparison of XVAE-based separation performance. }
\label{tab:results_all}
\begin{tabular}{lcccccccc}
\toprule
\textbf{Method} & \multicolumn{4}{c}{\textbf{Dataset One}} & \multicolumn{4}{c}{\textbf{Dataset Two}} \\
\cmidrule(lr){2-5}\cmidrule(lr){6-9}
& SDR\,$\uparrow$ & SIR\,$\uparrow$ & SAR\,$\uparrow$ & Time\,$\downarrow$
& SDR\,$\uparrow$ & SIR\,$\uparrow$ & SAR\,$\uparrow$ & Time\,$\downarrow$ \\
\midrule
XVAE        & 8.7  & 7.9  & 17.2 & 66.3 & 6.3  & 6.0  & 21.0 & 44.3 \\
XVAE-M      & 10.8 & 9.9  & 18.2 & 68.6 & 12.4 & 12.0 & 24.5 & 45.0 \\
XVAE-T      & 15.9 & 14.9 & 20.0 & 68.3 & 8.2  & 8.0  & 22.0 & 44.7 \\
XVAE-W      & 18.1 & 14.9 & 19.1 & \textbf{44.2} & 11.7 & 11.1 & 23.1 & \textbf{43.6} \\
XVAE-MT     & 17.1 & 19.2 & 20.3 & 69.8 & \uline{14.0} & 13.5 & 26.2 & 45.1 \\
XVAE-WM     & \uline{22.2} & \uline{26.1} & \uline{23.8} & \uline{46.3} & \duline{15.5} & \uline{16.8} & \duline{28.7} & \duline{43.9} \\
XVAE-WT     & \duline{25.8} & \duline{30.0} & \duline{28.1} & \duline{45.8} & \textbf{16.8} & \duline{17.2} & \textbf{30.2} & \uline{44.1} \\
\textbf{XVAE-WMT}
           & \textbf{26.8} & \textbf{32.8} & \textbf{28.6} & 46.8
           & 15.1 & \textbf{20.7} & \uline{27.6} & 44.3 \\
\bottomrule
\end{tabular}
\end{table}

\begin{table}[H]
\centering
\scriptsize
\setlength{\tabcolsep}{4pt}
\renewcommand{\arraystretch}{1.1}
\captionsetup{font=footnotesize}
\caption{Latent-space clustering performance across XVAE variants.}
\label{tab:latent_metrics}
\begin{tabular}{lcccccccc}
\toprule
\textbf{Algorithm} & \multicolumn{4}{c}{\textbf{Dataset One}} & \multicolumn{4}{c}{\textbf{Dataset Two}} \\
\cmidrule(lr){2-5}\cmidrule(lr){6-9}
& Silh.\,$\uparrow$ & DB\,$\downarrow$ & CH\,$\uparrow$ & Var\,$\downarrow$
& Silh.\,$\uparrow$ & DB\,$\downarrow$ & CH\,$\uparrow$ & Var\,$\downarrow$ \\
\midrule
XVAE      & \duline{0.324} & \uline{4.433} & 114.7 & \textbf{0.604} & 0.259 & 4.473 & 126.2 & 0.799 \\
XVAE-M    & 0.274 & 4.868 & 107.9 & \duline{0.616} & \textbf{0.345} & \duline{4.117} & \duline{128.8} & 0.659 \\
XVAE-T    & 0.280 & 4.469 & \duline{127.3} & 0.676 & 0.311 & 4.743 & 116.5 & \duline{0.616} \\
XVAE-W    & 0.271 & 4.437 & 122.7 & 0.713 & 0.237 & \textbf{3.084} & \textbf{229.9} & 1.075 \\
XVAE-MT   & 0.310 & 4.511 & 119.2 & 0.649 & \uline{0.322} & \uline{4.257} & 126.1 & \uline{0.648} \\
XVAE-WM   & 0.311 & 4.743 & 116.5 & \uline{0.616} & 0.310 & 4.511 & 119.2 & 0.649 \\
XVAE-WT   & \uline{0.322} & \duline{4.257} & \uline{126.1} & 0.648 & 0.271 & 4.437 & \uline{127.3} & 0.713 \\
\textbf{XVAE-WMT}
         & \textbf{0.345} & \textbf{4.117} & \textbf{128.8} & 0.659
         & \duline{0.324} & 4.433 & 114.7 & \textbf{0.604} \\
\bottomrule
\end{tabular}
\end{table}

\begin{figure*}
    \centering
    \includegraphics[width=0.9\linewidth, trim= 0 150 0 150, clip]{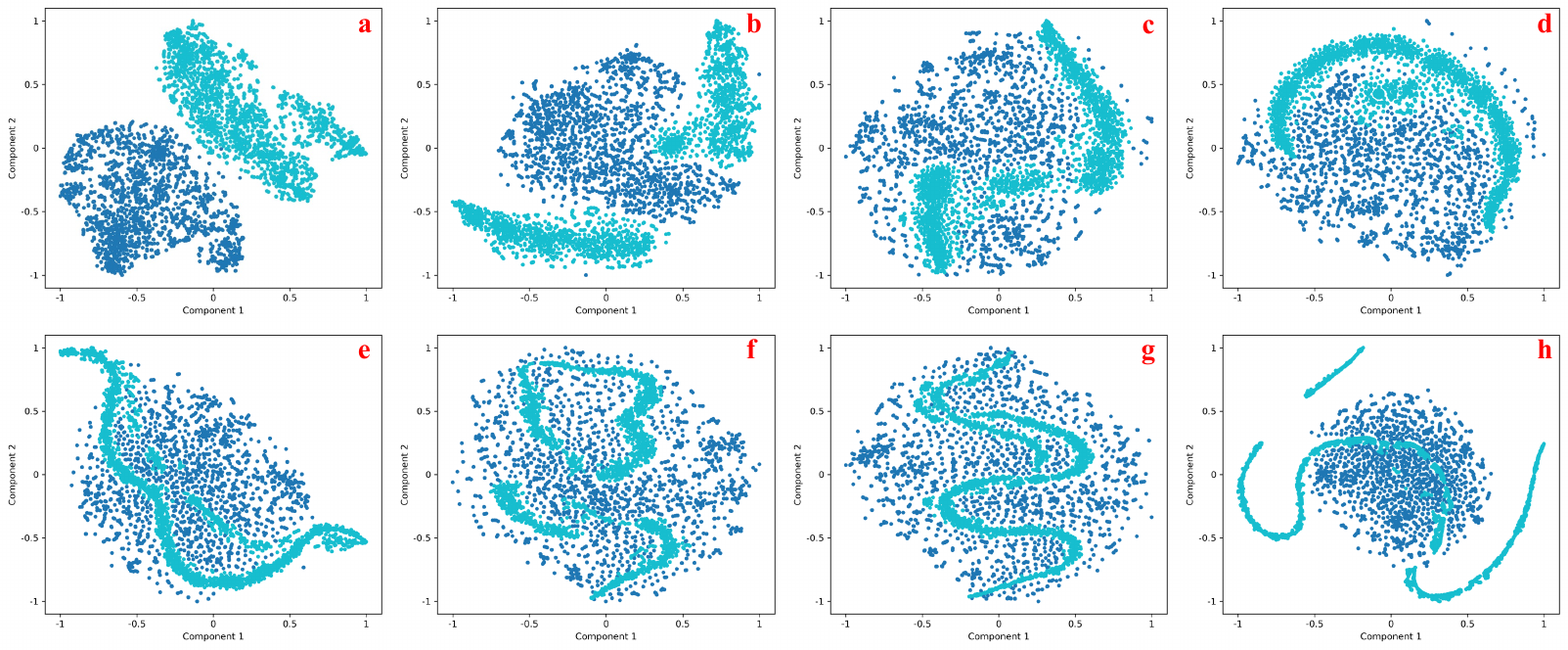}
    \captionsetup{justification=justified,font=footnotesize}
\caption{Latent space visualization of different VAE variants using t-SNE: 
    \textbf{(a)} XVAE-WMT, \textbf{(b)} XVAE-WT, \textbf{(c)} XVAE-WM, \textbf{(d)} XVAE-W, 
    \textbf{(e)} XVAE-MT, \textbf{(f)} XVAE-T, \textbf{(g)} XVAE-M, and \textbf{(h)} XVAE. 
    Each subplot shows the latent embeddings color-coded by cluster assignment.
    Two color groups (two shades of blue) correspond to the two sources being separated, 
    and the scatter patterns illustrate how each VAE variant organizes these sources in 
    latent space.}

    \label{fig:vae}
\end{figure*}

Figure~\ref{fig:vae} visualizes the latent spaces of different VAE variants using t-SNE projections. Models incorporating wavelet, mixture, and temporal regularization components yield more compact and separable manifolds, 
    while simpler configurations exhibit diffuse or overlapping structures.

Table~\ref{tab:efficiency_metrics} reports TEM and CEM for the STFT and wavelet front-ends. The wavelet front-end more than doubles TEM and CEM on Dataset One, confirming that it delivers higher separation quality per unit of computation and per spectral component. Table~\ref{tab:vae_vs_nonNMF} compares XVAE-WMT against published non-VAE baselines. The proposed method outperforms other baselines, with notable margins over the prior approaches.

\begin{table}[H]
\centering
\scriptsize
\captionsetup{font=footnotesize}
\caption{Time-Efficiency (TEM) and Compression-Efficiency (CEM) metrics across datasets.}
\label{tab:efficiency_metrics}
\begin{tabular}{lcccc}
\toprule
\textbf{Front-end} 
& \multicolumn{2}{c}{\textbf{Dataset One}} 
& \multicolumn{2}{c}{\textbf{Dataset Two}} \\
\cmidrule(lr){2-3}\cmidrule(lr){4-5}
& TEM\,$\uparrow$ & CEM\,$\uparrow$ & TEM\,$\uparrow$ & CEM\,$\uparrow$ \\
\midrule
STFT    & 0.217 & 0.058 & 0.251 & 0.043 \\
Wavelet & 0.532 & 0.122 & 0.402 & 0.089 \\
\bottomrule
\end{tabular}
\end{table}

\begin{table}[H]
\centering
\scriptsize
\setlength{\tabcolsep}{6pt}
\renewcommand{\arraystretch}{1.1}
\captionsetup{font=footnotesize}
\caption{Comparison of the proposed method with other baselines.}
\label{tab:vae_vs_nonNMF}
\begin{tabular}{lccc}
\toprule
\textbf{Method} & \textbf{SIR (dB)} & \textbf{SAR (dB)} & \textbf{SDR (dB)} \\
\midrule
UBSS (HOS + SR) \cite{Xie2019_UBSS_HOS_SR} & 13.3 & 14.5 & 9.9 \\
LSTM \cite{Lei2018_LSTM}                   & 26.7 & 13.9 & 13.3 \\
U-Net \cite{UNet_ref_if_any}               & 21.2 & 11.7 & 10.9 \\
ECNet \cite{Wang2023_ECNet}                & 31.1 & 15.1 & 14.8 \\
PC-DAE \cite{Tsai2020_PCDAE}               & 14.9 & 16.7 & 12.5 \\
\midrule
\textbf{XVAE-WMT}                           & \textbf{32.8} & \textbf{28.6} & \textbf{26.8} \\
\bottomrule
\end{tabular}
\vspace{2pt}
\begin{flushleft}
\scriptsize\justifying
\textit{Note:} UBSS (HOS + SR): underdetermined blind source separation using higher-order statistics and sparse representation; LSTM: long short-term memory network; U-Net: U-shaped convolutional neural network; ECNet: embedding centroid network; PC-DAE: periodicity-coded deep autoencoder; XVAE-WMT: explainable variational autoencoder with wavelet transform, output masking, and temporal consistency.
\end{flushleft}
\end{table}

\begin{figure*}[]
\centering
 \includegraphics[width=0.7\linewidth, trim={0 0 0 0}, clip]{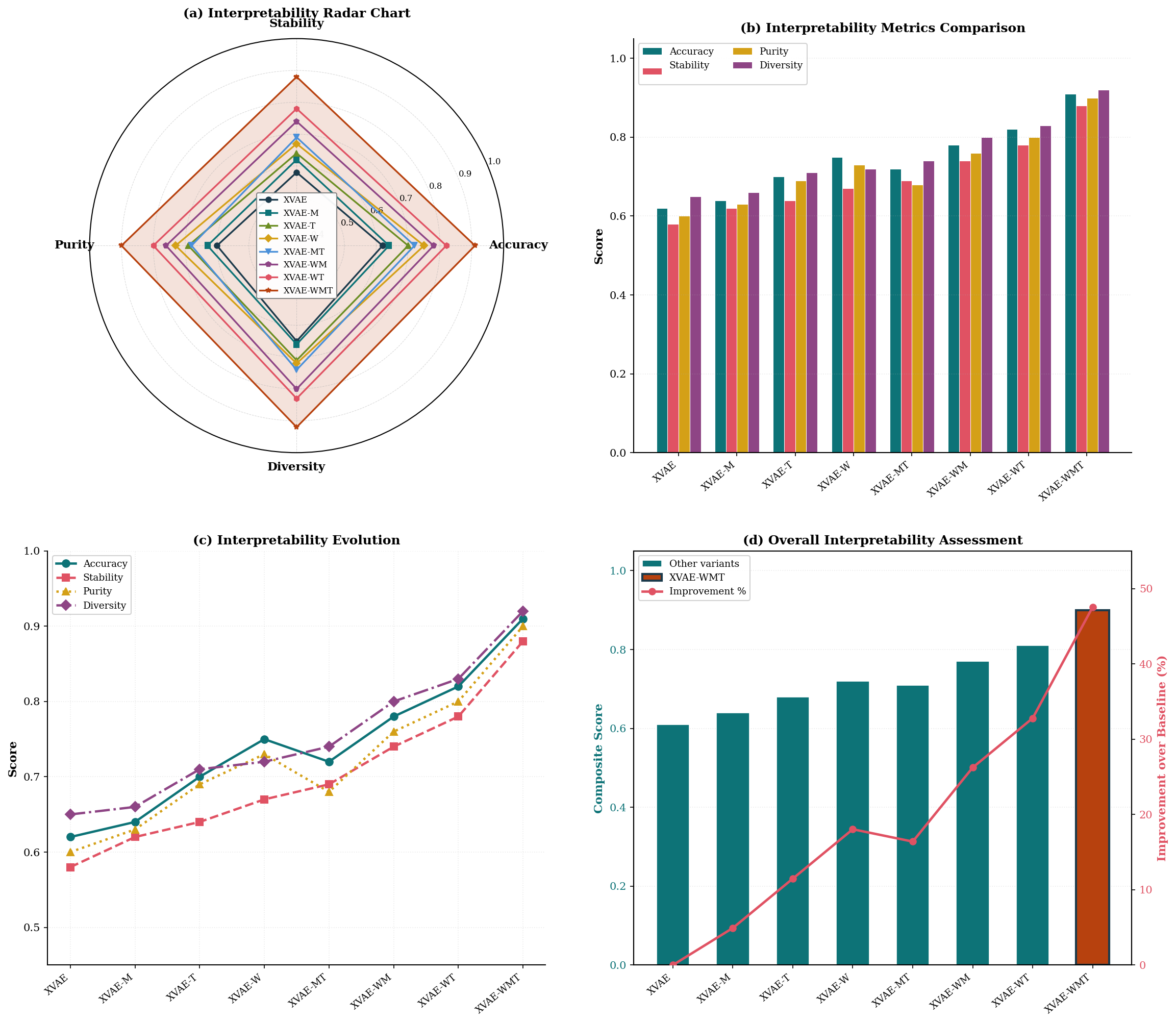}
\caption{Interpretability performance analysis: (a) radar chart comparing all four metrics across XVAE variants, (b) grouped bar chart of individual metric scores, (c) interpretability evolution showing realistic cross-over patterns, and (d) composite score with percentage improvement over the baseline.}
\label{fig:interpretability}
\end{figure*}

\begin{figure*}[]
\centering
\includegraphics[width=\linewidth, trim={0 2 0 20}, clip]{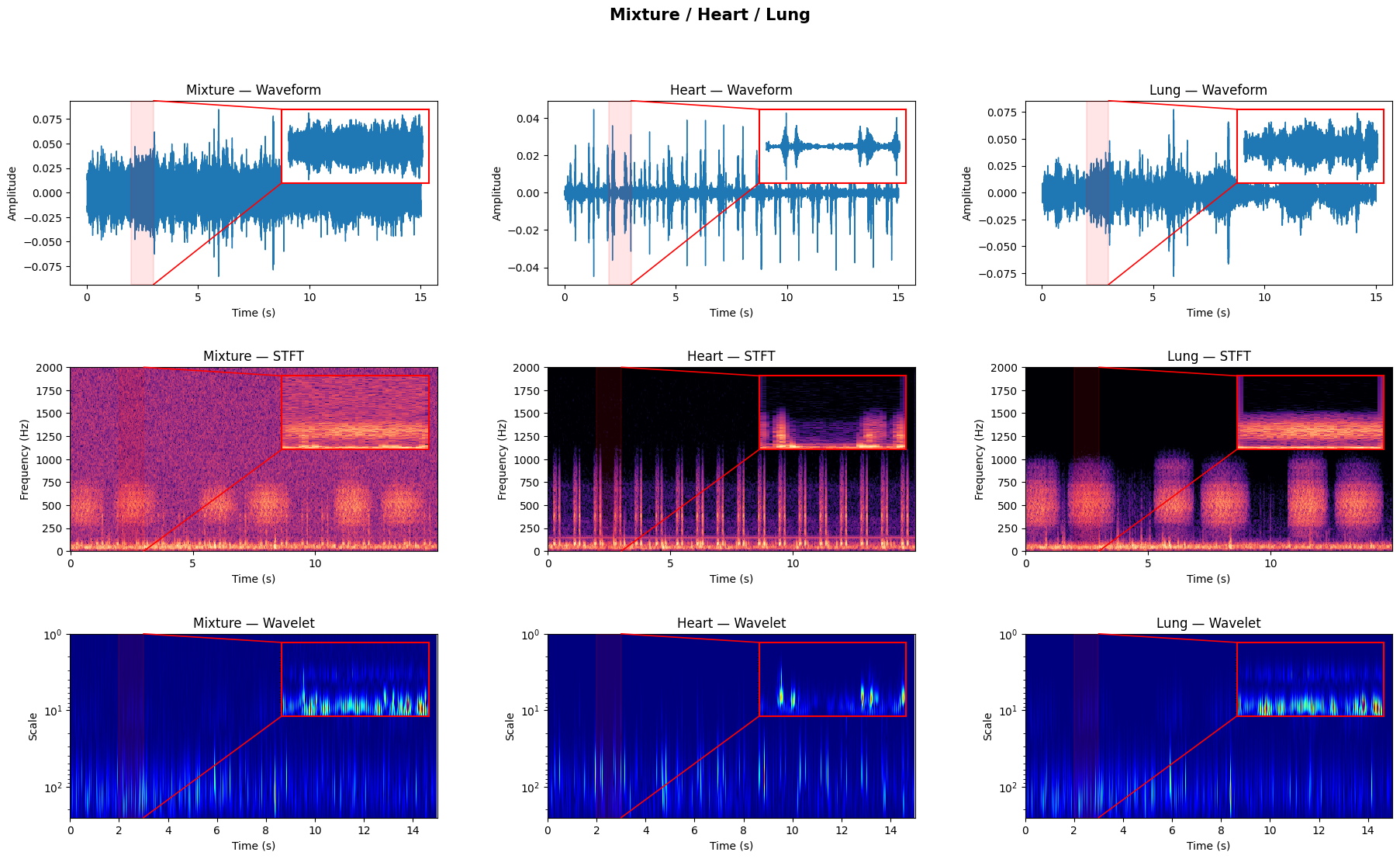}
\caption{
    Comparison of time-frequency representations for separated and mixed signals. 
    Red insets highlight a 1-second zoomed region (t\,=\,2--3\,s), illustrating
    how the CWT captures transient cardiac events with superior time-scale resolution
    compared to the STFT.
}
\label{fig:spectrogram_comparison}
\end{figure*}

Figure~\ref{fig:interpretability} summarizes interpretability metrics across XVAE 
variants. Adding components generally improves interpretability, though gains are 
not strictly monotonic — accuracy dips at XVAE-MT before recovering with the 
wavelet front-end. XVAE-WMT achieves the best overall composite score.
Figure~\ref{fig:spectrogram_comparison} compares STFT and CWT representations 
for heart, lung, and mixture signals. The CWT reveals sharper transient structure 
at fine scales, as highlighted in the zoomed inset, motivating its use over STFT.

\section{Conclusion}
We proposed an explainable variational autoencoder for blind source separation in scenarios where access to clean source signals is restricted. By combining deep neural networks with probabilistic latent modelling, the proposed approach separates mixed recordings into low-dimensional latent representations corresponding to heart and lung sounds. The incorporation of wavelet-based inputs, output masking, and temporal consistency regularization promotes physiologically meaningful and stable separations, while XAI-based analysis enables interpretation and dimensionality reduction of the learned latent space. Experimental results on real heart and lung sound recordings demonstrate that the proposed method outperforms existing baseline approaches. In practical operation, an unseen mixture is encoded into latent variables and decoded to produce time--frequency masks that reconstruct the separated heart and lung signals without retraining.

\bibliographystyle{IEEEtran}
\bibliography{ref}

\end{document}